# Two-photon polymerization of optical microresonators for precise pH sensing

Anton V. Saetchnikov*[1], Elina A. Tcherniavskaia[2], Vladimir A. Saetchnikov[3], and Andreas Ostendorf[1]

[1]*Chair of Applied Laser Technologies, Ruhr University Bochum, 44801 Bochum, Germany*
[2]*Physics Department, Belarusian State University, 220030 Minsk, Belarus*
[3]*Radio Physics Department, Belarusian State University, 220064 Minsk, Belarus*
*Correspondence to: Anton Saetchnikov: anton.saetchnikov@rub.de

**Abstract**

In water monitoring, environmental analysis, cell culture stability, and biomedical applications, precise pH control is demanded. Traditional methods like pH strips and meters have limitations: pH strips lack precision, while electrochemical meters, though more accurate, are fragile, prone to drift, and unsuitable for small volumes. In this paper, we propose a method for optical detection of the pH value based on the multiplexed sensor with 4D microcavities fabricated with two-photon polymerization. This approach employs pH-triggered reversible variations in microresonator geometry and integrates hundreds of dual optically coupled 4D microcavities to achieve the detection limit of 0.003 pH units. The proposed solution is a clear instance for the use-case oriented application of the two-photon polymerized structures of high optical quality. With benefits of the multiplexed imaging platform, dual 4D microresonators can be integrated alongside other microresonator types for pH-corrected biochemical studies.

**Keywords:** optical microresonator, two-photon polymerization, sensing, pH, whispering gallery modes, 4D printing

## Introduction

Various applications, including water quality monitoring, environmental analysis, biomedical, analytical chemistry, and pharmaceutical research, necessitate precise pH control. Two common laboratory methods for pH detection are ion colorimetry (using pH strips) and electrochemistry (using pH meters). Strips offer a quick and affordable option but lack precision, typically providing accuracy around one pH unit. Special strips designed for narrow pH ranges may feature color graduations as fine as 0.2 pH units and provide a response within several tens of seconds. Electrochemical pH meters are several orders of magnitude more precise, where the modern instruments with advanced electronics enables the resolution down to 0.001 pH units with response time down to several seconds [1]. However, they have drawbacks such as fragility, limited suitability for analyzing small volumes, challenges in miniaturization, signal drift without recalibration, sensitivity to temperature changes, and the need for regeneration before use [2]. Optical pH sensors are promising alternatives to traditional electrochemical methods with advantages like compactness down to submicrometer sizes, cost-effectiveness, and high sensitivity. Additionally, they are resistant to electromagnetic interference, support remote interrogation, can be minimally invasive, and eliminate the need for a separate reference sensor.

Optical pH detection utilizes changes in absorption or fluorescence in a solid matrix with encapsulated colorimetric or fluorescent indicator dyes, with fluorescence offering greater selectivity and sensitivity [3, 4]. Common performance of these pH sensors is defined by a limited dynamic range of $\pm$1.5 pH unit, response time of up to several minutes, and resolution of 0.01 pH unit. Another approach involves pH-triggered reversible expansion of specific polymers, optically detected by changes in refractive index. Both methods rely on establishing equilibrium within the sensitive material achieved through diffusion, with reduced layer thickness accelerating the response time. However, they are susceptible to temperature and ionic strength variations, and indicator dyes are additionally affected by photobleaching and ambient light. Unlike the indicator dyes, pH-sensitive materials also benefit from wider dynamic range with possibility to measure extreme pH values that are not accessible for electrochemical pH meters

[5]. Common refractive index-based pH detectors include fiber/waveguide [6], surface plasmon resonance (SPR) [7], and interferometric [8] schemes. Fiber/waveguide sensors are limited in performance by interaction length, SPR by rapid wave attenuation, and interferometric schemes offer high sensitivity at the cost of increased complexity. The most technologically advanced refractive index-based pH sensing schemes among them can be as accurate as 0.001 pH units with response times of few tens of seconds [5]. Optical microresonator-based sensing, with extended interaction length and other advantages, emerges as a competitive pH detection solution [9, 10].

In optical microresonators, electromagnetic field is confined within the dielectric material by the refractive index contrast, forming closed round loops [11, 12]. Whispering gallery mode (WGM) resonances, characterized by constructive interference of the waves within the circular cavity and low energy dissipation, exhibit high quality (Q-) factors [13]. Changes in resonator or environmental properties, such as temperature or biomolecule binding, display themselves as spectral changes in the WGM, enabling sensing [14–16]. These changes include resonance line shift, linewidth broadening, and modes splitting, along with variations in back-scattered field intensity caused by nearby nanoparticles [17, 18]. Various extensions, such as plasmonic-photonic schemes [19, 20], gain medium doping [21, 22], and exceptional point utilization [23], enhance sensitivity of the original WGM detection method [24]. Multiple optically coupled microresonators can transform symmetric WGM resonances into asymmetric Fano resonances with sharper line, improving sensing performance [25, 26]. Recently, deep learning algorithms paved the path to the affordable and robust sensing in the multiplexed detection scheme, allowing to monitor hundreds, or even thousands, of resonators simultaneously [27, 28].

Advancements in microresonator-based sensing technology have enabled the detection of various physical and chemical parameters [29–35]. For instance, pH detection has been achieved by coating microresonators with pH-sensitive layers and tracking the refractive index of the layer caused by its structural changes [36–38]. One method involves depositing polyelectrolyte (PAA/PAH) multilayers onto silica toroidal microresonators, ensuring detection at the level of 0.14 pH units [36]. Another approach functionalizes the interior surface of hollow bottle/bubble silica resonators [37, 38]. Unlike the previous example, this allows the main part of the electromagnetic field, instead of solely evanescent tail, to interact with pH-sensitive layer without significant loss in the loaded Q-factor. Hydrogel-embedded particles demonstrate the detectivity of 0.06 pH units and the response time down to 15 s [37]. Films of polyelectrolyte gels (PVA/PAA) sense pH in the range of 6-10, with suppressed reaction to refractive index changes [38]. Recently, pH variations in sulfactant solution were reported to be tracked via molecular reactions of Rhodamine B, used as the active medium to get WGM lasing [39]. This approach demonstrate step of 0.15, a response time of up to 60 s, and numerically estimated detection limit of 0.004 pH.

Thus, the main emphasis in the development of optical pH detectors has been given so far to providing a fast response within a wide pH range to become a valuable alternative to pH-meters. At the same time, alterations in pH at the level of 0.01 or below in a narrow range can have significant implications for analytical accuracy in biomedical research, clinical diagnostics, and environmental assessment. In particular, for studies of enzyme kinetics or immunoassays, where optical microresonators are effective sensing tools, maintaining a stable pH within a narrow range is critical to ensure reproducibility and accurate interpretation of the results. Here, the pH fluctuations in a buffered solution, where sensed components are immersed, can occur slowly over time as the buffer components react to external influences.

Using smart materials like hydrogels and polymers throughout the entire volume of the microresonator, rather than depositing them locally as a film on the cavity, amplifies the sensor spectral response. Fabrication of the structures capable to change function or shape in response to various external factors, such as temperature, humidity, light, electric field, pH, and chemical composition is known as 4D printing [40, 41]. 4D printing of micrometer scale structures having optical quality can be ensured by two-photon polymerization (2PP) [42]. We have recently demonstrated the capability of this technique for fabrication of the 4D microresonators enabling loaded quality factors up to $10^5$ in aqueous medium [43, 44].

In this paper, we report a novel approach for high precision pH value detection for aqueous solutions by utilizing a multiplexed sensor of 4D optical microresonators fabricated with two-photon polymerization. We discuss the design and manufacturing of multiplexed sensors with integrated structures out of dual optically coupled 4D microresonators for simultaneous interrogation. We evaluate the sensing performance for detecting small pH changes in aqueous environment near pH = 7.4 using multiplexed sensors with both single and dual 4D microresonators. It has been revealed that, compared to single 4D microresonators, the dual structures exhibit enhanced spectral shifts and increased distinction between spectral features in response to pH-induced cavity shape changes. We show that the pH states of aqueous media can be determined explicitly in steps of 0.01 (limited by external reference pH meter). Together with the estimated detection limit of 0.003, this ranks our approach among the most sensitive pH value detection schemes. By taking the advantage of a multiplexed imaging platform with up to thousands of optical microresonators integrated, dual 4D microresonators combined with other types of microresonators are able to provide accurate biochemical studies resistant to pH alterations.

## Results and discussion

**Dual 4D microresonator design and fabrication** In order to fabricate the dual microresonator sensing structure with two-photon polymerization, we propose the model with two identical toroidal cavities allocated at vicinity to each other to enable optical coupling between them Fig. 1.

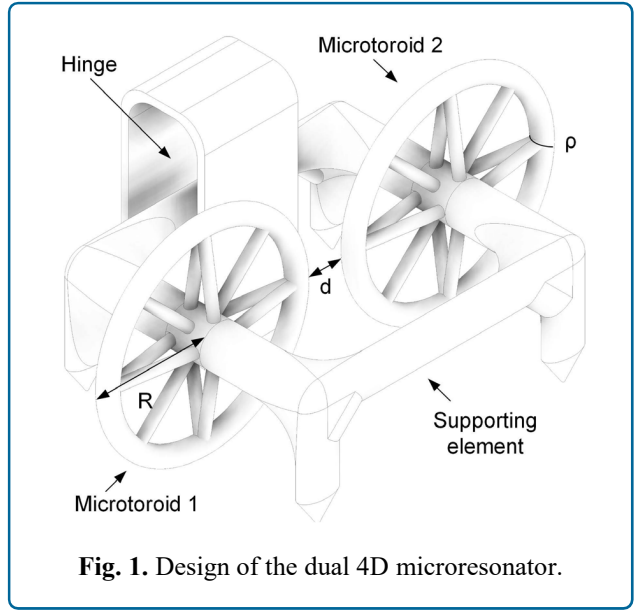

**Fig. 1.** Design of the dual 4D microresonator.

The features of the single 4D microcavity closely follow the parameters described in our prior publications [43, 44]. The microtoroid model is characterized by major and minor radii with values of $R = 21$ $\mu$m and $\rho = 1.8$ $\mu$m, respectively, which were optimized for minimization of the radiation losses along the curvature of the microcavity. The chosen microtoroid geometry allows the numerically estimated radiation limited Q-factor to approach the absorption limited one $(1.3 \times 10^7)$ [43]. Individual microtoroids are fixed to each other via supporting element for long-term stability and allocated at the distance, also gap, ($d$). The latter is searched in the range between 600 nm and 3 $\mu$m to enable the optical coupling between single microtoroids after fabrication. In order to mitigate the surface roughness and the associated scattering losses, primarily originating from the cross-linking between the polymer layers, as well as to precisely control the distance between the two resonators, a flexible support (hinge) is introduced to the model of the dual microcavity structure. Existence of this support enables the use of a layer-wise polymerization strategy, accommodating various orientations of the microresonator's symmetry axis during the polymerization and sensing phases. Reduction of scattering losses relies on optimizing the intersection between the polymerization voxels and the translation speed of the focus spot to ensure the smoothness of the microresonator rim. he upper limit for loaded Q-factor of microresonators fabricated with 2PP has been estimated at the level of $10^5$ in the water being constrained by the remaining surface roughness (Fig. 2). For particular dual 4D cavities the loaded Q-factor may drop down to $\lesssim 10^3$ due to the overlap of the modes of different orders excited in the microresonator. Scanning electron microscopy (SEM) image of a representative multiplexed sensor with dual 4D microresonators is represented in Fig. 3.

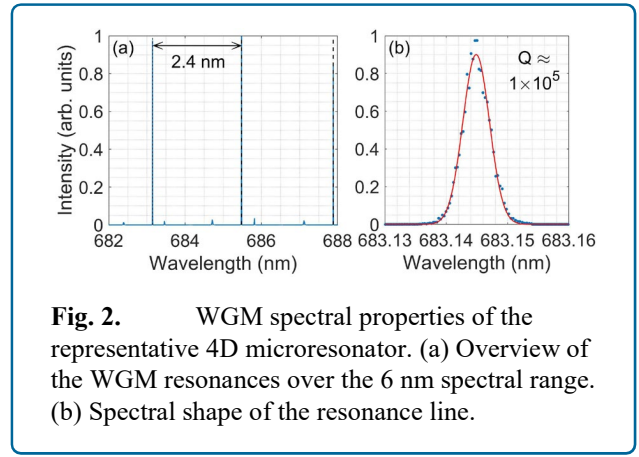

**Fig. 2.** WGM spectral properties of the representative 4D microresonator. (a) Overview of the WGM resonances over the 6 nm spectral range. (b) Spectral shape of the resonance line.

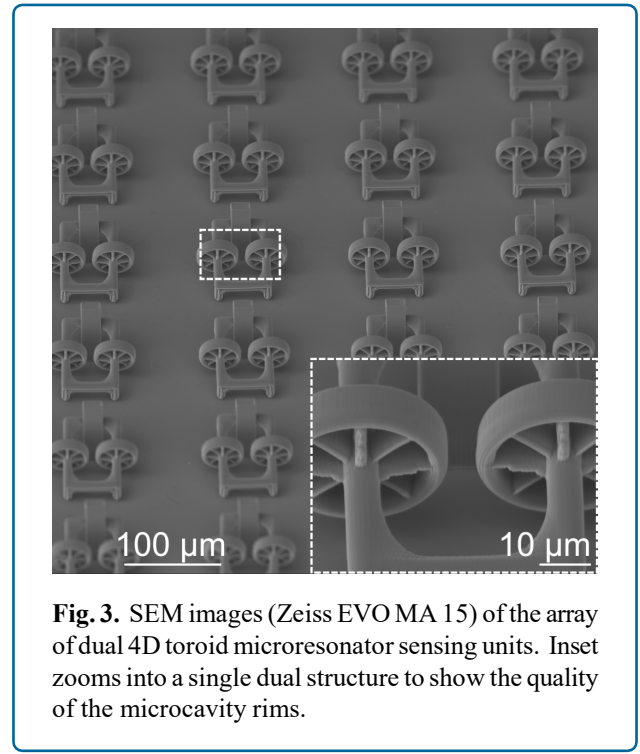

**Fig. 3.** SEM images (Zeiss EVO MA 15) of the array of dual 4D toroid microresonator sensing units. Inset zooms into a single dual structure to show the quality of the microcavity rims.

It was determined that for the applied photoresin illumination conditions the gap below 1.4 $\mu$m remains insufficient to avoid the cross-polymerization of two microresonators.

When the gap between toroids defined in the model approaches 2.2 μm, the microresonators stand apart on more than a wavelength. At this point the resonators can be already considered as two individual, optically independent, microcavities. The exemplary SEM images for the microresonators, which are polymerized together (direct contact), optically coupled, and optically independent, are shown in Fig. 4.

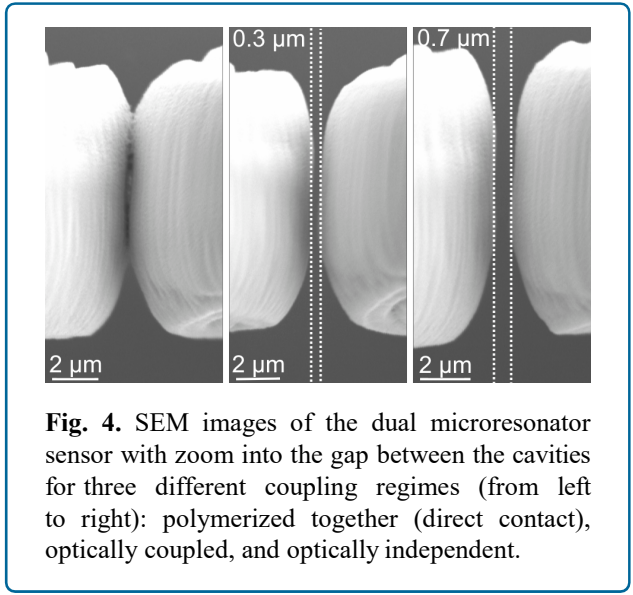

**Fig. 4.** SEM images of the dual microresonator sensor with zoom into the gap between the cavities for three different coupling regimes (from left to right): polymerized together (direct contact), optically coupled, and optically independent.

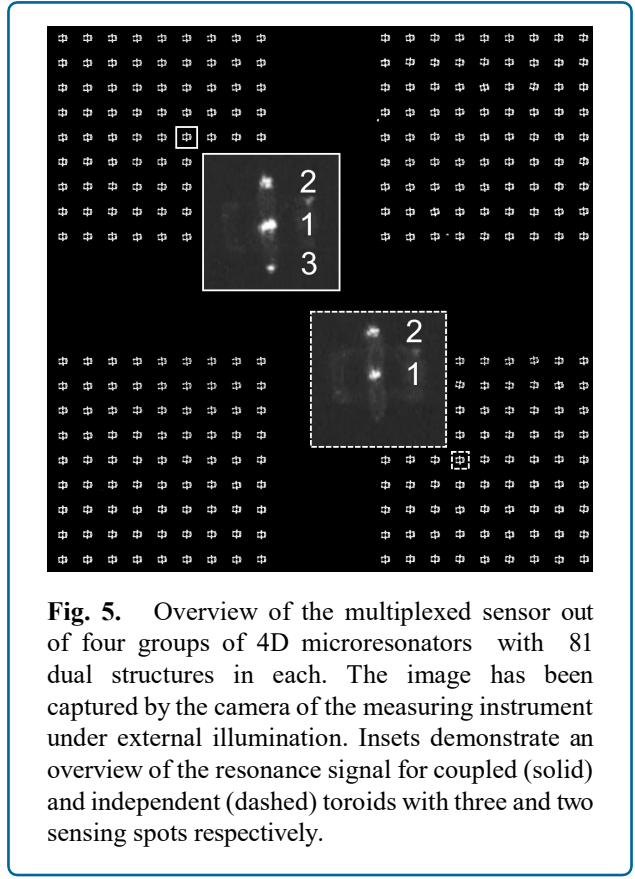

**Fig. 5.** Overview of the multiplexed sensor out of four groups of 4D microresonators with 81 dual structures in each. The image has been captured by the camera of the measuring instrument under external illumination. Insets demonstrate an overview of the resonance signal for coupled (solid) and independent (dashed) toroids with three and two sensing spots respectively.

Despite the high precision provided by the two-photon polymerization, when the photoresist is illuminated at medium to high focal spot translation speeds, the standard deviation for the gap in the double microresonator structure reaches a hundred nanometers for the same model. For this reason, an array with hundreds of dual structures made by the same model enables to evaluate the sensing performance of the coupled dual 4D microresonators. The sensor samples with two different models of dual structures with the design gap of 1.6 and 1.8 μm are fabricated, where the first group predominantly comprises coupled and in-contact resonators, whereas the second group contains coupled and independent toroids. Figure 5 depicts an image of the fabricated multiplexed sensor sample captured in the measuring setup. The represented sample features four groups of dual toroid microcavities (two upper groups -1.6 μm and two bottom ones -1.8 μm) allocated with a relative offset of 100 μm.

In the multiplexed microresonator imaging scheme the signal of individual microresonator is characterized by appearance of a single light radiation (sensing) spot [43]. It is localized at the resonator edge along the laser beam propagation path and indicates dominant mode propagation direction in the cavity. In the geometry of the sample depicted in Fig. 5, the excitation direction is oriented from bottom upwards and thus the sensing spot appears on the top end of the toroid's image. As demonstrated in the insets of Fig. 5, the signal of the dual microresonator sensor is characterized by at least two sensing spots (1,2 in the Fig. 5), one per each toroid in the structure. The third sensing spot (3 in the Fig. 5) in case of the independent resonators is absent or remains weak which indicates only the back-radiation on the surface roughness of the first microresonator. When the microresonators are fabricated closer to each other and the interval between them approaches the distance suitable for optical coupling, there is a mutual energy transfer between the first and the second resonators. Thus, the strength of the signal in the third point becomes significant and indicates the secondary back-radiated waves that propagate in the first toroid. The radiation spot related to the secondary waves inside the second toroid appears in close vicinity to the sensing spot of the first cavity and can not be resolved as a separate signal. Therefore, the third sensing spot is a clear indicator of mutual energy transfer between the toroids.

**PH detection with single 4D microresonator** The structures fabricated from SZ2080 material with two-photon polymerization exhibit the self-sensing capabilities enabling the realization of a 4D sensor. The self-sensing mechanism is a result of the expansion and contraction forces acting on the polymerized structure when immersed in different liquids. This phenomenon has been already revealed for single 4D microtoroids where their spectral shift response rises on the order of magnitude compared to the expected one for the pure evanescent field reaction [43, 44]. In case of significant distortions of the 4D resonator's shape, the spectral composition within the single free spectral range (FSR) may alter. Moreover, the natural diversity between the fabricated 4D structures results in a variance of the spectral composition for individual structures. Therefore, to address this, we introduced the second parameter, mode intensity, in addition to the spectral shift. This parameter is dedicated to describe the excitation efficiency for WGMs of different orders and has been estimated as the radiated light intensity integrated over the single FSR where the spectrum is corrected for shift in advance. At first, the sensing response of an array with single 4D microresonators for pH-value detection has been studied. Alterations of the pH value around pH = 7.4 are validated with external pH meter and ensured by mixing the phosphate buffer saline (PBS) solution with small portions of 0.2M NaOH (increase) and 3.7% HCl (reduction). PBS buffer is able to provide a stable pH environment, making it more suitable for validation of the sensing performance compared to, e.g., deionized water which, in turn, lacks buffering capacity and is susceptible to pH changes when exposed to external factors such as dissolved gases, atmospheric $CO_2$ absorption, and contamination. It is anticipated that variations in final viscosity among the adjusted PBS solutions are negligible.

The PBS buffer is known for maintaining the pH within a relatively narrow range, even with changes in temperature. Around the room temperature the observed shifts in pH due to variations of several °C are generally small, amounting to only a few hundredths or even thousandths of a pH unit. Here, with higher temperatures pH value tends to decrease. As a result, the changes in pH values within the measured solutions for the reported configurations can be considered as negligible. At the same time, the temperature variations in the sensing chamber affect the spectral properties of microresonators caused by thermo-optic and thermoelastic effects. This impact has been analyzed on example of the sensing phases representing the WGM signal of the microcavities in the pure PBS environment (which could last in total up to 30000 s for one single measurement). The spectral shift linearly follows the small temperature alterations with different negative slope values for particular 4D microresonator. Based on the determined temperature response for each microresonator, the experimental data has been adjusted for the temperature-related shift. An example of the temperature altered spectral shift response summarized over an array with single 4D microresonators is represented in Fig. 6.

Here, for the first 5000 s of the measurement the sensing chamber is filled with the pure PBS solution and for the next 5000 s with +0.02 pH altered PBS solution. As it can be seen from the graph, as long the temperature remains stable, the averaged spectral shift also remains constant. Then at 5000 s when the pH value is increased by 0.02 the averaged spectral shift rises by ≈10 pm and remains stable for the next 1000 s. Then, the temperature increases by 0.3 °C over the next 4000 s that leads to back-shift of the resonance wavelength to initial position measured for the pure PBS. Thus, the temperature response for this example can be estimated at the averaged level of ≈33 pm/°C. Considerable confidence interval for spectral shift demonstrated in Fig. 6 clearly indicates the dispersion of the sensing properties of the microresonators caused by the natural dispersion in fabrication of individual microcavities and their illumination. For this reason the selected parameters are first calculated for each sensor and then generalized by applying principal components analysis (PCA) for the whole group [27]. In contrast to the averaging among the microresonators, where each of them has equal contribution in the generalized response, PCA effectively captures the most significant variance within the dataset. For the generalized spectral shift, only the first principal component (PC1) has been considered, since it commonly reflects more than 80% of the shift variability for the array. For the generalized mode intensity, more than only PC1 has been accounted. This manner of representing the spectral variations offers an advantage of accommodating the varying sensitivities of the microresonators while mitigating the influence of localized fluctuations or noise, which are identified as insignificant components through the application of PCA.

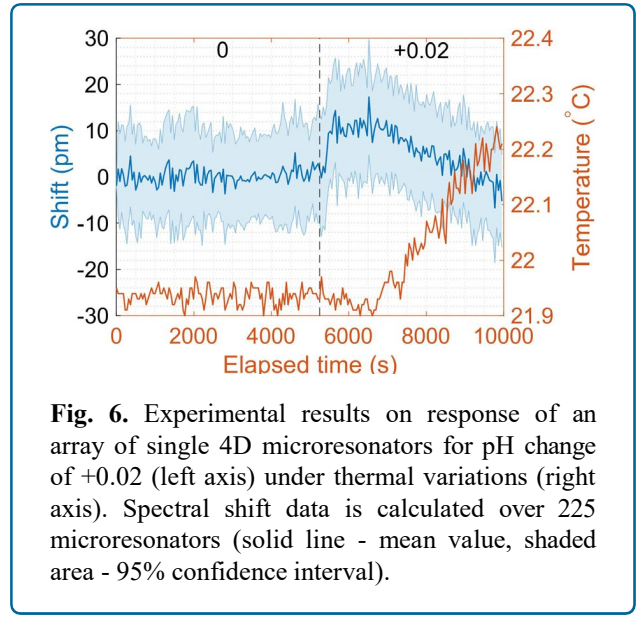

**Fig. 6.** Experimental results on response of an array of single 4D microresonators for pH change of +0.02 (left axis) under thermal variations (right axis). Spectral shift data is calculated over 225 microresonators (solid line - mean value, shaded area - 95% confidence interval).

The experimental results for the spectral responses of single-microtoroid 4D sensor are represented in Fig. 7. The results show clear spectral reaction of the multiplexed sensor with single 4D microresonators on pH value changes. Spectral responses as the generalized spectral shift for pure and NaOH-augmented PBS solutions (0, +0.07, +0.12, +0.2, +0.29 pH) is represented in Fig. 7(a). The sensor reaction on HCl augmented solutions (-0.05, -0.08, -0.12, -0.2 pH) are demonstrated in Fig. 7(b). Each solution was pumped for 6000 s through the measurement chamber followed by pumping of the PBS solution for 6000 s. As can be seen from the figures, the changes in pH value lead to the clear dynamics of the generalized spectral shift response, were the sign of the pH value alteration matches the direction of the resonance spectral shift. This indicates that 4D toroids swell, when the pH value increases, and shrink, when it reduces. Also, the greater is the difference in pH value, the stronger is the 4D microresonator response. It is also revealed that the saturation values of the generalized shift linearly depend on small variations of the pH value in both upward and downward directions. The dynamical variations represented in Fig. 7(a),(b) indicate that the time interval required to reach the equilibrium lasts longer when the difference between pH values of the sensed liquids increases and may exceed 1000 s already at the absolute pH change of 0.2. When further increasing the step in the pH value between two solutions measured in series up to the order of unity (the maximum pH value change of 5 has been measured), the steady state for the generalized spectral shift is not attained within the chosen measuring range of 6000 s.

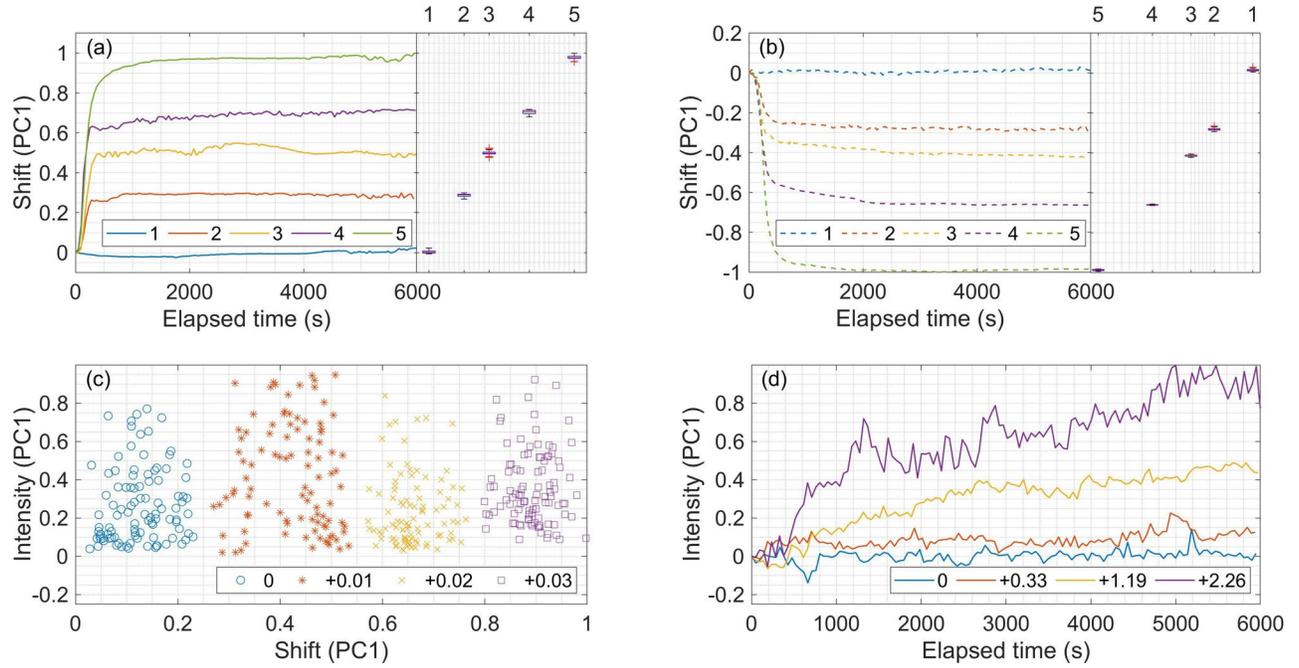

**Fig. 7.** Experimental results on pH-value detection with an array of single 4D microresonators. (a) Dynamics of the generalized spectral shift as a scaled first principal component (PC1) over 6000 s for (a) increasing [0 (1), +0.07 (2), +0.12 (3), +0.2 (4), +0.29 (5)] and (b) decreasing [0 (1), -0.05 (2), -0.08 (3), -0.12 (4), -0.2 (5)] pH values. (c) Allocation of pH-values with the step of 0.01 in the space of the generalized spectral shift/mode intensity. (d) Dynamics of the generalized mode intensity over 6000 s for significant pH-value alterations from +0.33 up to +2.26.

In order to estimate the pH value detection limit for the single 4D microresonators, tiny portion of NaOH solution is added to PBS to attain the step in pH value at the accuracy level of the reference pH meter (+0.01, +0.02 and +0.03). Each of these solutions has been pumped over 6000 s, as it has been done previously, and the saturation time points for the generalized spectral shift have been extracted. It has been determined that for the pH value changes below 0.03, the response time does not exceed 200 s. This is caused by refilling of the sensing chamber $\approx$ (300 $\mu l$) with incoming liquids at the selected pumping speed (100 $\mu l$/min). To verify whether the different pH values can be distinctly separated from each other, we allocate them in the space of the generalized spectral shift and mode intensity (see Fig. 7(c)). The results show that the pH states can be explicitly separated with the step of 0.02, where the cluster formation for the spectral data is guaranteed by the generalized spectral shift whereas the generalized mode intensity (PC1) spreads in the same region for all measured states. By increasing the pH value step by the order of magnitude (+0.33) the measured dynamics for the generalized mode intensity starts to exhibit a slight discrepancy from the baseline (Fig. 7(d)). When the pH value alteration approaches the unit order, the generalized mode intensity becomes valuable for detection of different environmental states. In case of big pH changes the 4D microresonator shape undergoes strong swelling/shrinkage, that eventually leads to a change in the excitation efficiency of WGMs of different orders within the coupling region.

**PH detection with dual 4D microresonator** In the next step, we focused on analyzing the spectral variations of the dual 4D microresonators induced by changes in the pH value. Similarly to the single 4D microresonators, swelling and shrinking of each toroid of the dual structure are initiated by varying the pH value. In turn, this leads to changes in the gap between the cavities and affects the conditions for coupling the modes between them. Here, the prominence of the individual WGMs in the spectrum rises or declines depending on the initial conditions. Due to the natural variance of the gap size between the toroids within an array, some of the dual 4D microresonators may not have the optical contact. For this reason, from the whole array on the sample only the structures with prominent signal in the back-radiated light spot (# 3 in Fig. 5), i.e., structures with signal appearing at more than three standard deviations relative to the noise level, have been included into analysis. Thus, the dual toroids that are in physical (direct) or optical contact are selected. With considerations above, we repeated the measurement at minimum increment (0.01) of the pH value for PBS solutions mixed with NaOH as it was previously done for single microresonators (Fig. 8).

The responses for the generalized mode intensity have been classified into two groups according to dynamics severity, where all signal spots (see Fig. 5) are considered as independent signals. The first group of signals are characterized by expressed variations in dynamics with pH value (Fig. 8(a)). Here, the saturation values rise with bigger step in the sensed pH level, which points out the intensified changes in spectral mode structure. Together with the generalized spectral shift, the mode intensity based responses for measured pH values states with the step of 0.01 form separate clusters that can be explicitly separated (Fig. 8(b)). Hence, when the optical contact is established, the multiplexed sensor with dual 4D microresonators demonstrates the improved at least twice sensing performance compared to the 4D sensors based on single toroids. Extra cluster (1*) with responses in pure PBS after measurement cycles of solutions with altered pH is added to denote the ability of the sensor to recover the baseline and thus its repeatability. Clusters indicating original sensor response and response after several measurement cycles have a large overlapping area with the mean offset of the scatter data cloud not exceeding the initial spread range. With this the baseline recovery after measurement of the small pH value alteration can be defined as sufficient. The theoretical detection limit for the dual 4D structures is estimated by considering the sensitivity for generalized mode intensity ($S$), using the formula $LOD = 3\sigma/S$ [45, 46], where $\sigma$ is standard deviation of the generalized mode intensity for the baseline. A linear model describes the relationship of the generalized mode intensity over small variations of the pH value and is used to calculate parameters $S$ and $\sigma$ mentioned above (Fig. 8(c)). It has been determined that alterations in pH value at the level of 0.003 can be effectively determined.

Another group of responses of dual sensors is characterized by random variations in the dynamics of the generalized mode intensity for all pH states studied. At the same time, the dynamics of the generalized spectral shift shows differences with partial overlaps of time points for different pH states. This group is expected to indicate the dual resonators with physical contact, where swelling of the microtoroid caused by the pH value rise does not lead to significant disturbances in conditions for coupling the modes between the microcavities.

In order to study the linkage between the mode structure variations of different sensing spots of the coupled resonators, the mode intensity dynamics are considered as independent signals and merged together for each dual structure. The principal components calculated from the mode intensity data for linked sensing spots of the dual sensor for the highest step of the pH value in the batch (+0.03) is shown in Fig. 8(d). The plot represents the dynamics of the first four PCs (together accounts for >60% of the variability in the sensing data) for three signal spots of the coupled cavities. It can be clearly seen that different PCs indicate various direction and expression of the dynamics for mode intensity among signals of the coupled resonators. Since PC is a linear combination of the original microresonator signals, its direction does not necessarily coincide with direction of the actual mode intensity changes. For this reason, only the mutual consistency of the temporal variations in terms of the sign was investigated. In this sense monitored PCs demonstrate all available combinations of spectral alterations, where PC1 describes consistency of the back-scattered signal (3) with the signal of the first resonator (1), PC2 with the second toroid (2), PC3 with neither first nor second cavity, and PC4 with both signals. Numbering of the sensing spots are the same as depicted in Fig. 5 and denotes the first resonator, the second one, and the back-radiated signals, respectively. In comparison to any of the first three PCs, the PC4 accounting less than 6% of the original data dispersion shows weak response to the measured pH alteration. Together with the same direction of the dynamics for all signals, this case is expected to describe the response caused by the changed coupling conditions between the cavities and the optical prism.

An overview of the spectral mode changes for initial (0) and increased pH value (+0.03) captured for the third sensing spot of two exemplary dual microresonator structures is represented in Fig. 9. The demonstrated spectra are limited by the single FSR, since its variations are estimated to be below the spectral resolution and consequently remain non-resolvable. The demonstrated spectra were acquired for each phase under saturation conditions and corrected for the spectral shift to denote spectral shape variations. For the first dual sensor (Fig. 9(a)) a clear energy redistribution between modes is observed, where suppression and promotion of separate modes within the single FSR occur. Besides suppression of the first mode as a result of the increasing pH level, its loaded Q-factor decreases on $\approx$ 15% which is linked to the changes in the coupling limited Q-factor. The second mode, being originally completely suppressed, appears as a distinct peak after increasing the pH value.

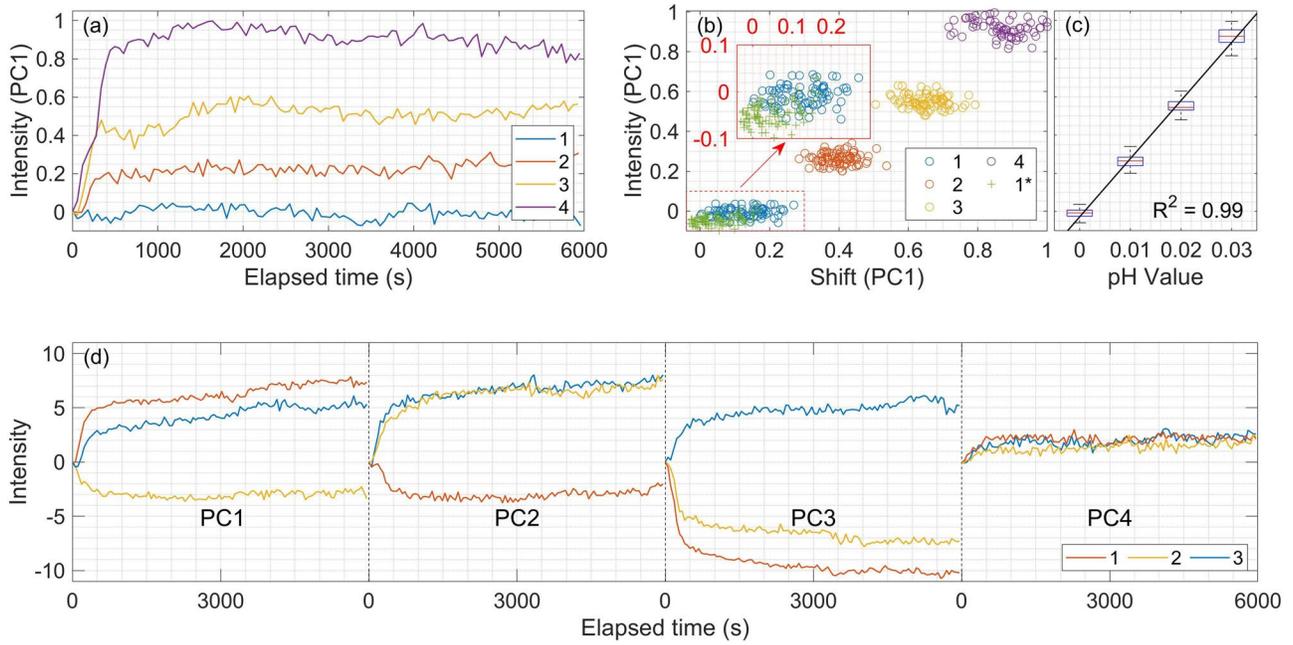

**Fig. 8.** The spectral response of the multiplexed sensor of optically coupled dual 4D microresonators on pH-value variations around pH = 7.4 measured for 6000 s. (a) Temporal dynamics of the generalized mode intensity (in form of the scaled PC1) for different pH values [0 (1), +0.01 (2), +0.02 (3), +0.03 (4)]. (b) Scatter plot for stationary sensing data of different pH values [0 (1), +0.01 (2), +0.02 (3), +0.03 (4)] along with pH state 0 (1*) measured after the sensing cycles in the space of the generalized spectral shift and mode intensity. (c) Linear fit of the generalized mode intensity to changes in pH value. (d) The linkage between radiations spots sensing dynamics (1-center, 2-upper, 3-bottom) of the dual microresonator structure for sensing phase +0.03 pH.

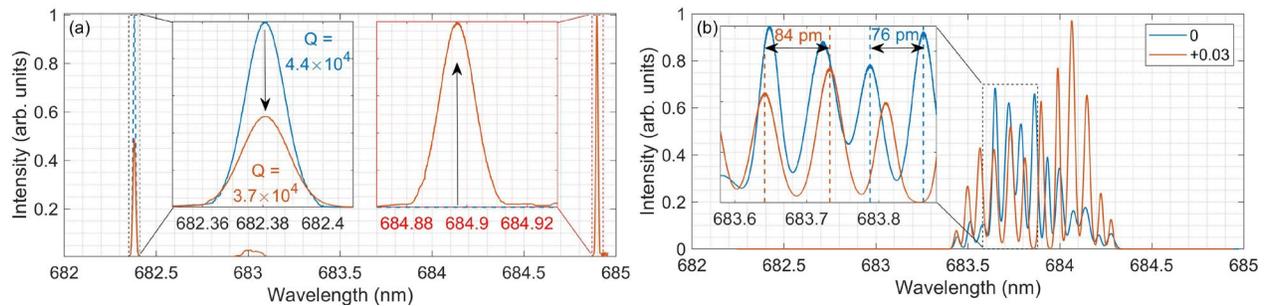

**Fig. 9.** Overview of the spectral changes within the single FSR for stationary states of different pH values (0 and +0.03) measured at the third sensing spot for two exemplary dual 4D microcavities.

The measured spectrum for the second exemplary sensor is characterized by the mode splitting phenomenon observable as the resonance lines allocated apart with a small interval (Fig. 9(b)). This response could be explained by the slight ellipticity of the path of the WGM propagation, where the split modes move away from each other with the deformation degree [47, 48]. As shown in (Fig. 9(b)), swelling of the microcavities that is induced by the increased pH value results in the increased pitching step from 76 to 84 pm which corresponds to the changes in the eccentricity level on just 0.3%.

As a result, the reported multiplexed sensor based on dual 4D microresonators shows the detection limit of 0.003 for pH changes that is determined from the difference in both spectral shift and mode intensity dynamics. However, the ultimate precision is possible when small alterations of the pH value are monitored. When increasing the step of pH value by one order of magnitude (compare to Fig. 8), the non-linear character of the spectral mode changes for optically coupled microresonators has to be considered. An example of the generalized mode intensity response for the pH value changing by -0.94, -0.51, +0.4, and +1.5 from the initial value of pH=7.4 is depicted in Fig.10.

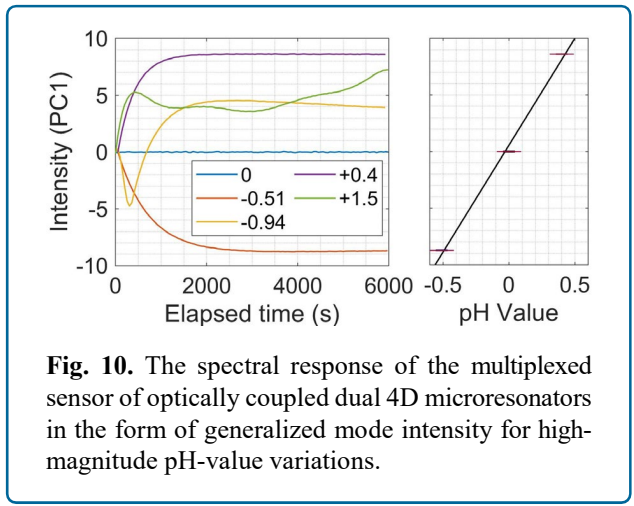

**Fig. 10.** The spectral response of the multiplexed sensor of optically coupled dual 4D microresonators in the form of generalized mode intensity for high-magnitude pH-value variations.

These results clearly show that different sign of the pH value alterations (-0.51 and +0.4) results in the opposite direction of dynamics for the generalized mode intensity similarly to the generalized spectral shift. When reducing the pH value by 0.94 or increasing it by 1.5, the complex response appears. This is most evident for pH value reduction by 0.51 where the generalized mode intensity moves towards negative values with the increased slope but shortly after the changes begin (at $\approx$300 s), the dynamics has a turning point. Such behavior indicates that at high alterations of pH value, the mode structure undergoes a series of significant spectral changes and each of them impacts non-linearly the whole spectrum. While the variations of pH value in the range of $\pm$0.5 are described by a linear function, different pH values separated by $\gtrsim$1 cannot be unambiguously distinguished by the generalized intensity of modes at saturation. For this reason, the high precision in measurement of the pH value with dual 4D microresonators for wide range should involve two steps: tracking the generalized spectral shift for the preliminary pH level estimation followed by its refinement via analysis of the generalized mode intensity. The moderate speed in response of the 4D microresonators to high pH alterations is caused by the diffusion that occurs over the entire microresonator volume. Thus, the 4D microresonator as a pH sensor is highly efficient for detection of small and slow environmental alterations and presents limited performance for tracking of rapid and significant pH value changes. However, taking into account the clear accordance of the spectral shift dynamics to the pH value, the method of dynamical responses analysis based on deep learning algorithms, as we previously proposed in [28], can be applied. Hereby, the decision time can be reduced multiple times compared to the time required to observe the stationary state, that we will reflect in the future studies.

Further complication of the coupled structures design for the multiplexed sensor construction may allow the introduction of preferential nature of mode changes and/or intensity variations in the back-scattered signal. Particularly realization of the dual microresonator structure with 4D cavities of different dimensions and variable optical linkage with external coupler will eliminate constraints on the narrow excitation bandwidth for the laser source, which, in turn, will further improve the affordability of the detection capabilities.

## Conclusions

In this paper we proposed a method for high-precision tracking of small pH alterations in the biochemical solutions. It is enabled by a multiplexed sensor with single or dual 4D optical microresonators produced via two-photon polymerization. The sensor features tight integration of many hundreds of similar structures for simultaneous tracking while individual sensing units are characterized by advanced post-fabrication material response. It exhibits the transformation of the spectral features triggered by the pH-induced cavity shape responses alongside the boosted spectral shift due to structural changes in individual 4D microresonators. Sensing performance of the proposed pH sensor has been studied for tracking opposite pH value alterations for biochemical buffer solution (pH = 7.4). We showed that the triggered structural changes of WGMs in optically coupled 4D microresonators have enabled the detection limit at the level of 0.003 pH unit. This sensing performance allows to allocate the multiplexed sensor with dual 4D microresonators among the most sensitive pH detection schemes currently available. Furthermore, owing to the remarkable integration capabilities with other types/functional groups of microresonators on the same substrate in an affordable manner within the multiplexing imaging platform, the proposed approach becomes a notable solution for multiparameter biochemical sensing.

## Materials and methods

**Chemicals** For the fabrication of the coupled dual microresonators we chose the sol-gel SZ2080 photoresin, incorporating the photoinitiator 4,4'-Bis-(diethylamino)-benzophenone and another monomer, 2-(dimethylamino) ethyl methacrylate (DMAEMA) [49]. Within this composition, the secondary monomer functions as a radical quencher throughout the polymerization. This modification enables to increase the spatial resolution of the polymerization process and realizes the diffusion-assisted two-photon polymerization that is chemical analogue of the Stimulated Emission Depletion Microscopy (STED) inspired multiphoton lithography. As a result, the polymer matrix of the fabricated structures is endowed with increased surface-to-volume ratios. These characteristics can be finely tuned by adjusting the illumination conditions and thus allowing the gap between the polymerized areas $\approx$ 100 nm or below.

To prepare the phosphate buffer saline solution, we utilized deionized water to obtain the concentration of 0.01 M phosphate buffer, 0.0027 M potassium chloride, and 0.137 M sodium chloride, maintaining the pH of 7.4 at the temperature of 25°C. Different levels for pH value for the test solutions have been achieved by adding small portions of either hydrochloric acid (HCl) to decrease pH, or sodium hydroxide (NaOH) to increase pH. All chemicals for preparing the solutions were obtained from Sigma-Aldrich.

**Fabrication process** Substrates for allocating the multiplexed sensing array of dual optically-coupled microresonators were cover glasses of 150 $\mu$m thickness. The substrates have been covered by $\approx$ 400 nm layer of low (water-matched) refractive index adhesive in advance to optimize the overlap between the evanescent fields of the coupling prism and microresonator and thus to enhance the coupling-limited Q-factor [44]. The layer uniformity has been ensured by the spin-coating procedure at 3000 rpm for 30 s. After adding a photosensitive material to pre-processed substrate, it is heated at maximum 80°C until the solvent is evaporated. Finally, the solidified droplet of the photosensitive material on a substrate is positioned within the two-photon polymerization (2PP) setup.

Fabrication of the dual microresonator sensors with 2PP is conducted using a custom-built optical setup. It incorporates a mode-locked Ti:Sa laser system (Tsunami, Spectra Physics) with emission wavelength of 780 nm, repetition rate of 82 MHz, and pulse duration of 90 fs. The setup is supplemented by acousto-optical modulator serving as a shutter, a galvo scanner that allows redirection of the laser beam within the plane parallel to the substrate, and linear stages enabling precise positioning of the sample relative to the galvo scanner. All together this enables the three-dimensional versatility in the photopolymerization process. The high resolution in fabricating the polymer structures with increased field of view is facilitated by focusing the laser light via 40x objective with NA of 0.95. The sample with the photoresin is layer-wise illuminated according to the predefined path for the laser spot calculated from the computer-aided design (CAD) model of the structure to be fabricated.

We have finely tuned the illumination parameters to expedite the polymerization process, so that the fabrication of hundreds of dual microresonator structures could be enabled in an acceptable time slot while maintaining the smoothness and regularity of the cavity rims. The laser power has been fixed at 32 mW; the slicing and hatching distances were set at 200 nm and 100 nm, respectively; the laser spot translation speed was selected at 100 mm/s for the rims and at 200 mm/s for the supporting components. With above-mentioned illumination configuration, the production time for a dual microresonator structure does not exceed three minutes. To eliminate the non-polymerized remains, the samples underwent the wet-chemical treatment lasting for 20 minutes in a 4-methylpentan-2-one developer, followed by a 10-minute immersion in 2-propanol. By necessity, circulation of the developing solutions was performed for improved remains removal in the gap between the cavities. Finally, the substrates with microresonators were left to dry for several hours, allowing the solvent to dissipate. Then, the samples were cleaned with UV-ozone procedure for 5 min in order to improve the wettability of the microresonator surfaces and ensure the penetration of the sensed liquids between the coupled cavities.

**Measuring instrument** Signal collection from an array with both single and dual microresonators has been realized in the multiplexed microresonator imaging scheme developed earlier [27, 32, 44]. The resonances in the cavities are excited in the optical prism based scheme. Here, unlike the typical photodiode-based monitoring of the excitation channel, the signal radiated by each individual microresonator is captured in the far field using a camera. By choosing the cavities with the axis of rotational symmetry oriented parallel to the optical prism, the signals from many microresonators can be simultaneously collected. In order to ensure same excitation conditions for all cavities, the microresonator array is illuminated with a collimated beam. The area for the effective microresonator interrogation is restricted to the diameter of $\approx$ 8 mm that is defined by the achromatic optical collimation package (60FC-T-4-M40-24, Schaefter+Kirchhoff). In order to address the elongation of the collimated beam projection in the propagation direction at the prism excitation surface, an anamorphic prism pair is introduced to alter the laser beam profile from circular to the elliptical one. The light source is a diode laser (Velocity, New Focus) with tunable emission wavelength from 680 to 690 nm and the linewidth of 200 kHz. The laser beam profile is governed by propagation through a single-mode fiber (630HP, Thorlabs). The polarization of the beam has been tuned by a polarization controller (FPC030, Thorlabs) to excite TE modes in the microcavities, and the wavelength was monitored by a wavemeter (WS7-30, HighFinesse). The use of the monochrome low noise camera (pco.edge 4.2, PCO) allocated in the far field as a signal detector enables spatial separation of signals from different structures as well as separation between the signals of the first, the second resonators, and the back-scattered signal within each dual resonator structure. The fluctuations in temperature of the liquids being sensed are controlled by embedded temperature sensor (PT 100) located within the sensing head. For reference pH measurements of solutions, a calibrated pH meter (Greisinger GMH 5530) is used to ensure a measurement accuracy at the level of 0.01.

The substrate with an array of dual microresonators is allocated on the optical prism via the immersion oil to ensure stable coupling conditions. The assembly of the sensing head involves the combination of the prism holder, optical prism, sensor sample, and flow chamber components, which are then carefully positioned on a precise three-axis stage. The parameters of the fluid flow are set by the pressure-based controller (LINEUP FLOW EZ, Fluigent) in conjunction with a flow rate sensor (FLOW UNIT, Fluigent). The instrument allows the sequential pumping of fluids from different containers selected by valve (M-SWITCH, Fluigent) through the array of dual microcavities. The fluid pumping rate is kept constant at 100 $\mu$l/min which reduces the possible impact of the viscosity of the liquids onto pH response of the 4D microresonator sensor.

**Acknowledgement**
We thank the group of Dr. Maria Farsari (IESL-FORTH) and, particularly, Dr. Gordon Zyla for providing the photoresin. The authors Andreas Ostendorf and Anton Saetchnikov are grateful to the German Federal Ministry for Research and Education (BMBF) for partly funding this work under the VIP+-Programme in the project IntellOSS, 03VP08220. Furthermore, the authors thank Parisa Farsari for her help in fabrication and SEM characterization of the samples.

**Conflict of interest**
The authors declare no conflict of interest.